\documentclass[twocolumn,showpacs,preprintnumbers,amsmath,amssymb]{revtex4}

\usepackage{graphicx}
\usepackage{dcolumn}
\usepackage{bm}
\usepackage{umlaute}

\begin{document}


\title{Precision measurement of the metastable $^3{\mathrm{P}}_2$ lifetime of neon
}

\author{Martin Zinner}
\author{Peter Spoden}
\author{Tobias Kraemer}
\author{Gerhard Birkl\footnote{Author to whom correspondence should be addressed.}}
\author{Wolfgang Ertmer}

\affiliation{%
Institut f\"ur Quantenoptik, Universit\"at Hannover, Welfengarten 1, D-30167 Hannover
}%

\date{\today}

\begin{abstract}

The lifetime of the metastable $^3{\mathrm{P}}_2$ state of neon
has been 
determined
to $14.70(13)\, \mathrm{s}$ (decay rate $0.06801(62)\, \mathrm{s}^{-1}$)
by measuring the decay in fluorescence of an ensemble of
$^{20}$Ne atoms trapped in a magneto-optical trap (MOT).
Due to low background gas pressure ($p < 5 \times 10^{-11} {\mbox{mbar}}$)
and low relative excitation to the $^3{\mathrm{D}}_3$ state
(0.5 \%
excitation probability) operation only small corrections
have to be included in the lifetime extrapolation.
Together with a careful analysis of residual loss mechanisms in the MOT
a lifetime determination to high precision is achieved.
\end{abstract}

\pacs{32.70.Cs, 32.70.Fw, 32.80.Pj}
\maketitle

In contrast to its importance for various active fields of research
covering such a wide range as
atomic physics, quantum optics, and nuclear physics,
there still has been no
measurement of the natural lifetime of the metastable
$^3{\mathrm{P}}_2$ state \cite{LS} of neon with sufficient precision.
A selection of research activities
profiting from an improved measurement
include the quest for
Bose-Einstein condensation (BEC) of metastable neon,
advanced atomic structure calculations, the investigation
of ultracold collisions, and even precision tests of the
electroweak theory being currently persued by
investigating the nuclear decay of an optically trapped sample of 
$^{19}$Ne in the $^3{\mathrm{P}}_2$ state
\cite{berkeley}.

Ongoing research directed towards
BEC of metastable neon atoms
in the $^3{\mathrm{P}}_2$ state \cite{Beijerinck,uns} clearly will
benefit from an accurate knowledge of the state's lifetime.
This includes the optimization of the production process as well as
the study of collision processes such as Penning-ionization \cite{Penning} 
and elastic s-wave scattering.
Exciting new physics 
complementing the work on metastable helium \cite{HeBEC1,HeBEC2}
can be expected for  
a $^3{\mathrm{P}}_2$ neon condensate.
We would like to point out
the study of higher-order correlations \cite{Correlations} and 
the intriguing possibility of a modification of the
$^3{\mathrm{P}}_2$ decay rate due to the high phase-space
density or the phase coherence in a BEC \cite{Savage}.

Significant advance in atomic structure
calculations has already been and will be
further triggered \cite{Kim} by the work presented here.
To date only a preliminary experimental value of 22 s for the
$^3{\mathrm{P}}_2$ lifetime of neon exists \cite{ShimizuNeon}
with no detailed investigations having been performed \cite{Shimizuprivate}.
A precise determination of the lifetime will
put a more stringent test on theory.
For most rare-gas atoms, the $^3{\mathrm{P}}_2$ lifetime depends sensitively on 
electron correlations and relativistic corrections.
The latter have only a minor effect
for the case of neon, making neon specifically interesting
for a critcal test of  electron correlations.
In addition, our measurement will
close the chain of $^3{\mathrm{P}}_2$ lifetime measurements for rare-gas atoms
beyond helium performed in recent years
for Ar ($38_{-5}^{+8}\, \mathrm{s}$ \cite{ShimizuArKr}),
Kr ($39_{-4}^{+5}\, \mathrm{s}$ \cite{ShimizuArKr} and $28.3_{-1.8}^{+1.8}\, \mathrm{s}$ \cite{Walhout}),
and Xe ($42.9_{-0.9}^{+0.9}\, \mathrm{s}$ and $42.4_{-1.3}^{+1.3}\, \mathrm{s}$ \cite{Rolston}), respectively.
This will deliver the input for
a detailed systematic investigation of the Z-dependence of
metastable lifetimes.


\begin{figure}[b]
\centering
    \includegraphics[width=200pt, keepaspectratio]{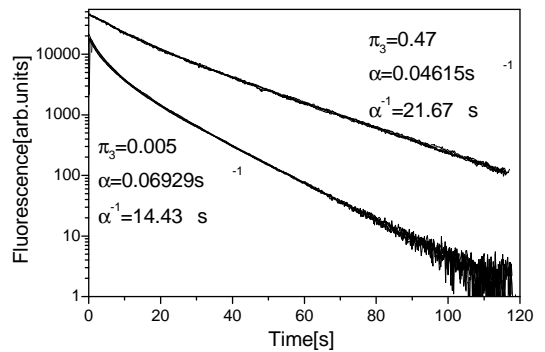}
\caption{Decay of the fluorescence for a $^{20}$Ne ensemble in a MOT
for the given excitations to the $^3{\mathrm{D}}_3$ state of
$\pi_3=0.005$ ($\Delta/\Gamma=-5.3$, $I/I_0=1.75$, 
$p = 3.4 \times 10^{-11} {\mbox{mbar}}$) and
$\pi_3 = 0.47$ ($\Delta/\Gamma=-0.16$, $I/I_0=24$,
$p = 4.7 \times 10^{-11} {\mbox{mbar}}$).
Both sets of data 
consist of five curves each, demonstrating
the
high reproducibility of our decay measurements.
}
\label{fig:curves}
\end{figure}


An accurate experimental determination of
the $^3{\mathrm{P}}_2$ lifetimes of rare-gas atoms
other than helium has only been achievable by the ability
to prepare cold atom samples in a MOT
\cite{ShimizuNeon, ShimizuArKr, Rolston, Walhout}.
For a precision lifetime measurement,
the coupling of the $^3{\mathrm{P}}_2$ to the $^3{\mathrm{D}}_3$ state
by the MOT light 
has to be considered
since it will modify
the observed decay rate.
To date, two different methods making use of a
variable MOT-on/off duty cycle have been employed,
either by directly measuring the rate of UV photons produced 
by the decay during a MOT-off period \cite{Rolston,Walhout}
or by extrapolating the observed decay rate for varying MOT-on/off duty ratios to vanishing
MOT-on periods \cite{ShimizuArKr}.

In our approach, we record the decay of fluorescence 
 in a
MOT (Fig.~\ref{fig:curves}) for different steady-state values of the population $\pi_3$
of the $^3{\mathrm{D}}_3$ state ($0.005 \!<\! \pi_3 \!<\! 0.47$). This
allows the extrapolation of the observed decay rate to $\pi_3 =
0$ thus giving the $^3{\mathrm{P}}_2$ lifetime. 
Extreme care has been taken
to achieve optimized experimental conditions such as low background
gas pressure ($p < 5 \times 10^{-11} {\mbox{mbar}}$) and 
MOT operation for such low values as $\pi_3 = 0.005$. This
minimizes the amount of required corrections dramatically
and the accuracy to which the  $^3{\mathrm{P}}_2$ lifetime of
Ne can be determined is improved considerably.

With no observable changes in shape and temperature of the atom sample, 
the temporal evolution of the observed fluorescence is proportional to
the temporal evolution of the number of trapped atoms $N\!(t)$.
The decay can be described by the differential equation
\begin{equation} \label{decay}
\frac{\mbox{d}}{\mbox{d}t}N\!(t)=-\alpha N\!(t) - \beta\! \int \!\! n^2\!(r,t) \, \mbox{d}^3\!r + \mathcal{O}(n^3).
\end{equation}
The experimental fluorescence data is fitted to a solution of this
equation. No absolute calibration of the atom number is required
for the determination of the decay constant $\alpha$. In
Fig.~\ref{fig:curves}, non-exponential two-body decay 
due to intra-trap collisions can be seen during the first 30 seconds.
Two-body losses depend on the integral over the
local number density $n(r,t)$ and are described by the parameter
$\beta$, the absolute determination of which
would require an absolute calibration
of $n(r,t)$, which is not the scope of this work.
An uncalibrated value of $\beta$, however, is determined in our
fitting procedure.
For all data runs
no contributions of order $n^3$ or higher could be observed.


\begin{figure}[b]
\centering
    \includegraphics[width=150pt, keepaspectratio]{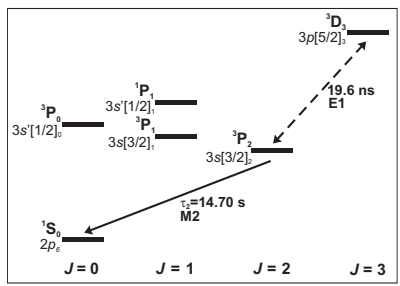}
\caption{Detail of the level scheme of neon. Atomic levels are labeled according to 
LS- and Racah notation.
Transitions are labeled by their multipolar character.
}
\label{fig:scheme}
\end{figure}


At low atom numbers
the decay is dominated by one-body losses with rate
%
%
%
\begin{equation} \label{alpha}
\alpha = \frac{1}{\tau_2} \left( 1 - \pi_3 \right)  + \frac{1}{\tau_3} \pi_3  + \gamma p +\mathcal{ML}.
\end{equation}
The parameter $\pi_3$ gives the population in the
$^3{\mathrm{D}}_3$-state. Thus $(1-\pi_3)N$ atoms decay from the
$^3{\mathrm{P}}_2$-state with lifetime $\tau_2$.
The combined
rate for all decay channels from
the $^3{\mathrm{D}}_3$-state other than the transition to the
$^3{\mathrm{P}}_2$-state (see Fig.~\ref{fig:scheme})
is given by $\tau_3^{-1}$. Background
collisions contribute with a rate of $\gamma p$ \cite{MOT1} with $p$ being the
pressure in the vacuum chamber and $\gamma$ being a constant
depending on the atomic and molecular species present.
The parameter $\mathcal{ML}$ summarizes possible MOT losses.
The main objective of this work is to determine the lifetime
$\tau_2$ from the observed decay rates $\alpha$ by extrapolating
to vanishing excitation ($\pi_3=0$), background pressure
($p=0$), and MOT losses ($\mathcal{ML} = 0$).
Achieving optimized starting conditions (small $\pi_3$ and $p$)
and minimizing 
potential contributions of $\mathcal{ML}$ by a systematic analysis
of the MOT characteristics have been essential.

\begin{figure}[b]
\centering
    \includegraphics[width=200pt, keepaspectratio]{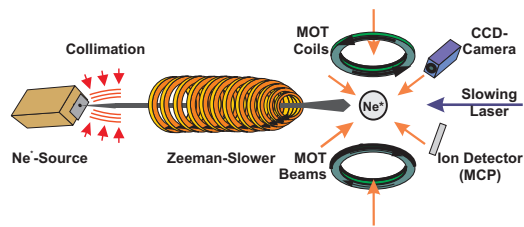}
\caption{Experimental setup.}
\label{fig:setup}
\end{figure}


Following \cite{MOT2}, the population $\pi_3$
is determined by:
\begin{equation}
\pi_3 = \frac{1}{2} \,\frac{C \cdot I/I_0}{1 + C \cdot I/I_0 + \left(  2 \Delta / \Gamma \right)^2 }.
\end{equation}
with 
the total intensity $I$ of the MOT light fields,
the saturation intensity $I_0=\pi h c \Gamma / 3 \lambda^3=4.08\, \mathrm{mW/cm}^2$
(linewidth $\Gamma= 2 \pi \! \times \! 8.18\, \mathrm{MHz}$,
$\lambda = 640.4 \,\mathrm{nm}$) and detuning $\Delta$.
The phenomenological parameter $C$
accounts for the deviation of the effective total intensity in a three-dimensional MOT from a
one-dimensional two-level system \cite{MOT2}.
We adopt the notion of \cite{Beijerinck}
and take $C=0.7$ with an uncertainty of $0.3$.
The intensity $I$ is
measured with an uncertainty
of 10 \%, 
the detuning $\Delta$
with an absolute uncertainty of $1\, \mathrm{MHz}$.
Therefore, for all data presented here the uncertainty in $C$ is
the dominant contribution to the uncertainty in $\pi_3$.

The experimental setup is shown in Fig.~\ref{fig:setup}.
A collimated beam of me\-ta\-stable atoms is decelerated
in a Zeeman-slower
with the slowing laser beam passing through the center of the MOT.
The MOT light field consists of six individual beams (diameter
$22\,\mbox{mm}$) with spatially filtered beam profiles.
All light fields are derived from a dye laser
which is frequency stabilized
to a neon RF-discharge.

For each
 run, the MOT is loaded for
$400\,\mathrm{ms}$ reaching a number of up to $4 \! \times \!
10^8$ atoms, a maximum collision-limited density on the order of $1 \!
\times \! 10^9 \,\mathrm{cm}^{-3}$
and a typical temperature below 1 mK. Then, the loading process is
terminated by blocking the atomic beam and the slowing laser beam
by mechanical shutters, and by switching off the magnetic field of
the Zeeman slower.
In the following measurement period of two minutes, the fluorescence of the trapped atoms is
recorded by a CCD camera
capturing five image frames per second.
%
%
Fluorescence decay curves
are obtained by integrating the counts of each frame over the region where atoms are present.
Great care has been taken to eliminate the influence of residual stray light.
The decay curves span up to four orders of magnitude in signal.
Fig.~\ref{fig:curves} shows two sets of data taken for different values of $\pi_3$.
For $\pi_3=0.005$ a decay rate of $\alpha = 0.06929(48) \, \mathrm{s}^{-1}$ and a respective
decay time of $\alpha^{-1} = 14.43 \, \mathrm{s}$ is observed.


\begin{figure}[b]
    \includegraphics[width=100pt, keepaspectratio]{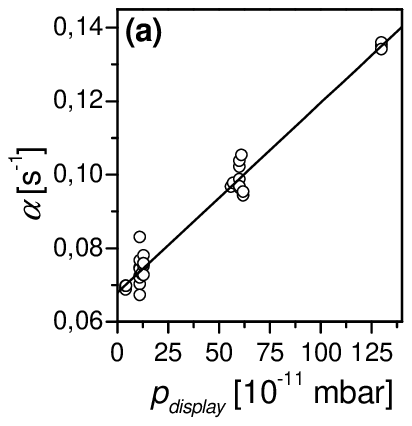}
    \includegraphics[width=100pt, keepaspectratio]{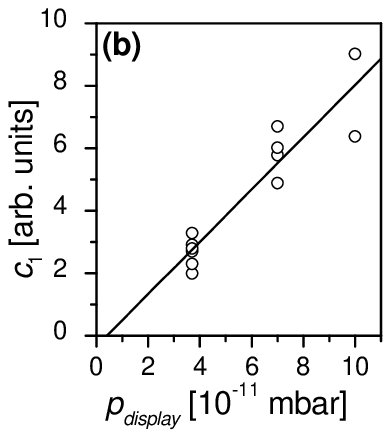}
\caption{(a)~Decay rates $\alpha$ obtained for
$\pi_3<0.1$ after extrapolation to 
$\pi_3 = 0$ as a function of the nominal background pressure $p_{display}$.
(b)~Coefficients $c_1$
proportional to the number of ionizing background collisions as a function of 
$p_{display}$ allowing the 
determination of the vacuum gauge offset. }
\label{fig:background}
\end{figure}


%
%
%
%
%
%
%
In the following sections we describe our procedure to extrapolate the observed decay rate $\alpha$
to the $^3{\mathrm{P}}_2$ decay rate $\tau_2^{-1}$.
For the extrapolation of $\alpha$ to vanishing background pressure
a systematic investigation of the MOT decay as a function of pressure
has been carried out
(Fig.~\ref{fig:background}(a)).
The pressure is given as
the nominal reading of the vacuum gauge (Balzers IKR070).
The specific composition of
the background gas is not known. However, we could observe a linear dependence of $\alpha$
on pressure over a range of almost two orders of magnitude,
indicating a pressure-independent composition of
the background gas.
From a linear fit for data taken for $\pi_3 < 0.1$, we can determine the slope
$\gamma = 5.2(1) \!\times\! 10^{7} \,\mathrm{mbar}^{-1} \,\mathrm{s}^{-1}$.
As a special feature of our setup,
we can directly determine a possible offset by simultaneously measuring the
fluorescence and the rate of ions
produced by collisions
using a multichannel plate (MCP) ion detector.
Under typical operating conditions
we detect ions of different origin:
(a) Ne$^+$-ions originating from intra-trap collisions
(rate
$\propto \int \!\! n^2\!(r,t) \, \mbox{d}^3\!r$).
(b) Ions which originate from collisions of
metastable neon atoms with residual gas constituents
(rate $\propto N(t) \times p$). 
Thus, the ion count rate $R_{\mathrm{ion}}(t)$ can be modelled by
%
%
\begin{equation}
R_{\mathrm{ion}}(t) = c_1 N\!(t) + c_2 \! \int \!\! n^2\!(r,t) \, \mbox{d}^3\!r,
\end{equation}
where $c_1$ is assumed to be proportional to $p$.
The values of $c_1$ obtained for different pressures are shown in
Fig.~\ref{fig:background}(b). From a linear fit we can determine
the offset of the pressure gauge reading to $(4 \pm 7) \times
10^{-12} \,{\mathrm{mbar}}$ also allowing for a vanishing
offset. By combining the results for $\gamma$ and the
offset, we can extrapolate the observed
decay rates to $p =0$. Due to the already low
background pressure for our decay measurements only small
corrections arise: At $p = 5 \!\times\! 10^{-11}
\,{\mathrm{mbar}}$ the background collision rate of $\gamma p =
1/(420\,{\mathrm{s}})$ is more than a factor of 25 smaller than
the observed decay rate $\alpha$.



\begin{figure}[b]
\centering
    \includegraphics[width=120pt, keepaspectratio]{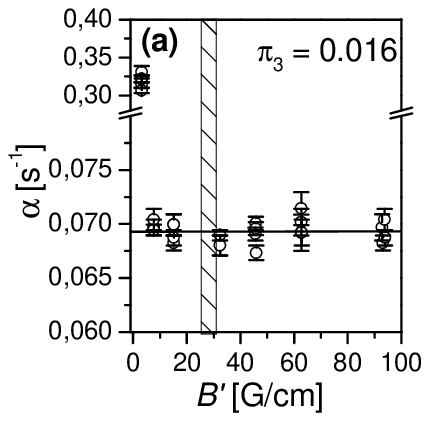}
    \includegraphics[width=120pt, keepaspectratio]{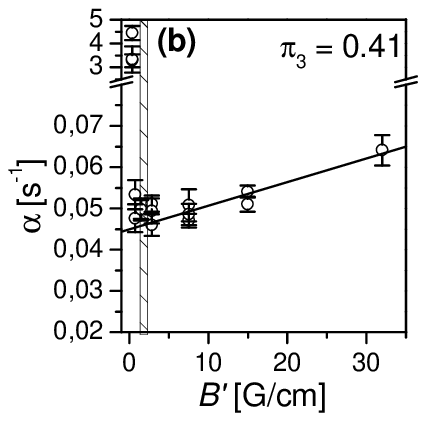}
\caption{Systematic investigation of MOT losses. Decay rates are plotted 
as a function of the magnetic field gradient
for two different excitations (a) $\pi_3=0.016$ and (b)
$\pi_3=0.42$. Shaded areas indicate the gradients
of typical operating conditions.}
\label{fig:gradient}
\end{figure}


We also investigated the
possibility of additional atom losses from the MOT
(contribution $\mathcal{ML}$
in Eq.~(\ref{alpha}))
attributed to a finite trap depth $U_0$ or finite escape velocity $v_e$
\cite{Weiner}.
We assume 
that a lossless MOT would result from an infinite trapping volume
leading to infinite $U_0$ and $v_e$. 
In principle, this could be achieved 
for the applied magnetic field gradient $B'$
approaching zero if the MOT beams could be made infinitely large.
In all practical implementations
the trap depth will remain finite due to the
finite size of the laser beams.
%
%
%
%
Thus, for a
given beam geometry, a variation in $B'$ should
allow to investigate the MOT stability and 
to quantify the MOT losses $\mathcal{ML}$
which
include diffusive motion out of the trapping volume \cite{Willems}
leading to an exponentially decreasing loss rate with increasing trap depth
and non-ionizing background gas collisions \cite{MOT1} 
 already being accounted for by the $p \to 0$ extrapolation.
The influence of the magnetic field gradient on the measured decay rate $\alpha$ is shown
in Fig.~\ref{fig:gradient}
for two values of $\pi_3=0.016$ and $0.41$, respectively.

In the most relevant case of small $\pi_3$ (Fig.~\ref{fig:gradient}(a)), 
for sufficiently large
$B'> 10\,\mathrm{G/cm}$ a variation of $B'$ does affect the measured $\alpha$
not more than
the uncertainties obtained from the fit.
Thus we conclude that for our operating regime (shaded region in Fig.~\ref{fig:gradient}(a))
the trap depth $U_0$ and escape velocity $v_e$
are sufficiently large and that
potential MOT loss contributions to the decay rate $\alpha$ 
are not significant.
For large $\pi_3$ (Fig.~\ref{fig:gradient}(b)), 
a variation of the MOT stability with $B'$ is observed.
Unfortunately, no quantitative model of the MOT at high saturation exists,
the application of the loss models discussed above is not straightforward,
and only a minimum and a maximum estimate of $\mathcal{ML}$ can be given:
Since our decay data for large $\pi_3$ have been taken at
$B'=1.5\,\mathrm{G/cm}$ which gives the best MOT stability, it is
not clear whether MOT losses have a significant contribution on
the determination of $\alpha$ at all.
On the other hand, a linear extrapolation of $\alpha$
to $B'=0$
should overestimate the corrections to be made \cite{Willems,MOT1}
giving $\mathcal{ML}=0.005(2)\,\mathrm{s}^{-1}$ as an upper limit.
Losses for large $\pi_3$ only influence the
$\pi_3 \to 0$ extrapolation which
we take into account 
by incorporating $0\,\mathrm{s}^{-1}\!<\!\mathcal{ML}\!<\!0.005\,\mathrm{s}^{-1}$ 
into the uncertainties of 
this extrapolation.

\begin{figure}[t]
\centering
    \includegraphics[width=200pt, keepaspectratio]{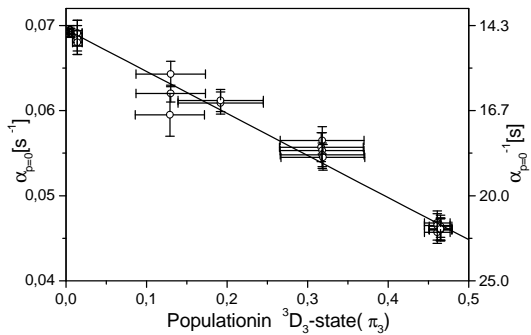}
\caption{Decay rates $\alpha_{p=0}$ 
as a function of the excitation
$\pi_3$ allowing the determination of $\tau_2$ in the 
limit of $\pi_3 \to 0$.}
 \label{fig:Alpha}
\end{figure}

After this systematic study of the influence of background
collisions and MOT losses
the determination of the $^3\mathrm{P}_2$ lifetime $\tau_2$ can be
obtained by 
extrapolating the measured decay rates to vanishing population $\pi_3$.
Only data for
a pressure below $5 \times 10^{-11} \, {\mbox{mbar}}$ are considered
and an extrapolation to $p=0$ is performed.
The dependence of $\alpha_{p=0}$ on $\pi_3$ is shown in
Fig.~\ref{fig:Alpha} exhibiting a simple linear behavior.
Displayed uncertainties in $\alpha_{p=0}$
are
given by the combined uncertainties in the fit of $\alpha$ and
uncertainties in the $p\!\to\!0$ extrapolation.
To minimize uncertainties 
in the  $\pi_3 \! \to \! 0$ extrapolation,
only data sets for $\pi_3 < 0.05$
and $\pi_3 > 0.4$ showing the smallest absolute uncertainties in $\pi_3$ have been included.
The final value of the $^3\mathrm{P}_2$ lifetime and decay rate are
$\tau_2=14.70(13)\,\mathrm{s}$ and $\tau_2^{-1}=0.06801(62)\,\mathrm{s}$,
respectively .
Table~\ref{tab:errors} gives a summary of contributions and
uncertainties for these values.
Since stable MOT conditions are directly accessible for
$\pi_3 = 0.005$ and $p < 5 \times 10^{-11}\, {\mbox{mbar}}$
only
unprecedented small corrections
are required. The final value for $\tau_2$ only deviates from the
uncorrected value of $\alpha^{-1} = 14.43 \, \mathrm{s}$ by 2 \%.
\begin{table}[t]
\caption{\label{tab:errors}Contributions to the
$\tau_2=14.70(13)\mathrm{s}$ extrapolation.}
\begin{ruledtabular}
\begin{tabular}{lrr}
MOT decay at $\pi_3=0.005$      &$0.06929\,{\mathrm{s}}^{-1}$   &$\pm 0.00048\,{\mathrm{s}}^{-1}$ \\
$\pi_3\to0$-extrapolation       &$+0.00028\,{\mathrm{s}}^{-1}$  & $\pm 0.00013\,{\mathrm{s}}^{-1}$  \\
$\mathrm{p}\to0$-extrapolation      &$-0.00156\,{\mathrm{s}}^{-1}$  &$\pm 0.00037\,{\mathrm{s}}^{-1}$   \\
\hline
$1/\tau_2$      &$0.06801\,{\mathrm{s}}^{-1}$   &$\pm 0.00062\,{\mathrm{s}}^{-1} $\\
\end{tabular}
\end{ruledtabular}
\end{table}

In addition to this result, extrapolating $\alpha_{p=0}$ to $\pi_3=1$, that is to a
hypothetical total population transfer to the
$^3{\mathrm{D}}_3$-state, is also possible. 
From this, we gain the first experimentally determined lower limit 
of $\tau_3\geq59\,\mathrm{s}$
for 
the combined rate $\tau_3^{-1}$ of decay of the
$^3{\mathrm{D}}_3$-state via all channels except the direct
decay to the $^3{\mathrm{P}}_2$ state.
However, this value depends sensitively
on the $\mathcal{ML}$-correction applied for large $\pi_3$
and only the stated lower limit is accessible.


%

To conclude we would like to compare our result
of $\tau_2=14.70(13)\,\mathrm{s}$ to
theoretical values obtained by different atomic structure calculations.
Prior to our work, published theoretical values  
were given in  \cite{Small} to 24.4 s, which can be
rescaled to 20.0 s using more accurate
input parameters,
and in \cite{Kim2} to 22 s.
Significant remaining discrepancies 
initiated a reexamination of atomic structure calculations for the whole set 
of $^3{\mathrm{P}}_2$ lifetimes of rare-gas atoms.
A preliminary result based on a multiconfiguration Dirac-Fock calculation
gives a value of 18.9 s for neon \cite{Kim}. A high degree of sensitivity
to the included electron correlations has been found.
Theoretical values are still expected to be larger than the
experimental value, since the inclusion of additional electron correlations
should further reduce the calculated lifetime.
Refined calculations are forthcoming \cite{Kim}.
Most recent ab-initio calculations using MCHF methods
give a lifetime of $16.9\,\mathrm{s}$ \cite{FroeseFischer}.
Considering the combined progress in experiment and theory, we anticipate
that the determination of the $^3{\mathrm{P}}_2$ lifetime of neon may continue to serve as 
an important test for precision atomic structure calculations in the future. 
%

We thank Y.-K. Kim for valuable discussions and are grateful for financial support by the DFG within the
{\it Schwerpunktprogramm SPP 1116}.

\bibliography{apssamp}

\begin{references}


\bibitem{LS} The notation of states throughout this paper is based on
LS-coupling. Racah notation
is included in Fig.~\ref{fig:scheme}.

\bibitem{berkeley} see e.g. \texttt{http://weak0.physics.berkeley.edu/weakint/}\\
\texttt{research/neon/}.

\bibitem{Beijerinck} S.J.M.~Kuppens et al., 
Phys. Rev. A {\bf65}, 023410 (2002).

\bibitem{uns} M.~Zinner, Ph.D. thesis, unpublished (2002);
\texttt{http://www.iqo.uni-hannover.de/html/ertmer/}\\
\texttt{atom\_optics/nebec/nebec.html}.

\bibitem{Penning} F.M.~Penning, Naturwissenschaften {\bf15}, 818 (1927);

\bibitem{HeBEC1} A.~Robert et al.,
Science {\bf292}, 461 (2001).

\bibitem{HeBEC2} F.~Pereira Dos Santos et al.,
Phys. Rev. Lett. {\bf86}, 3459 (2001).

\bibitem{Correlations} M.~Yasuda, F.~Shimizu,
Phys. Rev. Lett. {\bf77}, 3090 (1996);
E.A.~Burt et al.,
Phys. Rev. Lett. {\bf79}, 337 (1997);
W.~Ketterle, H.J.~Miesner,
Phys. Rev. A {\bf56}, 3291 (1997).

\bibitem{Savage} J.J.~Hope, C.M.~Savage,
Phys. Rev. A {\bf 54}, 3177 (1996).

\bibitem{Kim} J.P.~Desclaux, P.~Indelicato, and Y.-K.~Kim, private communication (2002).


\bibitem{ShimizuNeon} F.~Shimizu, Laser Spectroscopy IX, Academic Press, p. 444 ff, (1989).

\bibitem{Shimizuprivate} F.~Shimizu, private communication.

\bibitem{ShimizuArKr} H.~Katori, F.~Shimizu, Phys. Rev. Lett. {\bf70}, 3545 (1993).

\bibitem{Walhout} J.~Lefers et al.,
Phys. Rev. A {\bf 66}, 012507 (2002).

\bibitem{Rolston} M.~Walhout, A.~Witte, and S.L.~Rolston, Phys. Rev. Lett. {\bf72}, 2843 (1994).

\bibitem{MOT1} A.M.~Steane, M.~Chowdhury, and C.J.~Foot, J. Opt. Soc. Am. B {\bf6}, 2142 (1992).

\bibitem{MOT2} C.G.~Townsend et al.,
Phys. Rev. A {\bf 52}, 1423 (1995).

\bibitem{Weiner} J.~Weiner et al.,
 Rev. Mod. Phys. {\bf71}, 1 (1999).


\bibitem{Willems} P.A.~Willems et al.,
 Phys. Rev. Lett. {\bf78}, 1660 (1997).

\bibitem{Small} N.E.~Small-Warren, L.-Y.~Chow-Chiu, Phys. Rev. A {\bf11}, 1777 (1975).

\bibitem{Kim2} J.P.~Desclaux, P.~Indelicato, Y.-K.~Kim, cited in \cite{Rolston}.











\bibitem{FroeseFischer} G.~Tachiev,~C.~Froese~Fischer (2002), \texttt{http://atoms.vuse.vanderbilt.edu}.


\end{references}

\end{document}